# Integration of Spectral CT with PET and SPECT: Bringing Tissue Composition Information to Molecular Imaging

Guobao Wang, Kris Thielemans, Peter B. Noël, Joel S. Karp, Paul E. Kinahan

***Abstract*—Molecular imaging has been transformed by integrating positron emission tomography (PET) or single-photon emission computed tomography (SPECT) with x-ray computed tomography (CT). However, x-ray CT in hybrid imaging is used primarily for anatomical localization and for attenuation and scatter correction of emission data. Its ability to characterize tissue composition remains underutilized. Spectral CT, including dual-energy CT (DECT) and photon-counting CT (PCCT), can provide material-specific information, such as images of iodine concentration, bone or calcium fractions, electron-density, effective atomic number and other spatially-varying material properties. These capabilities create an opportunity to bring tissue-composition information into molecular imaging. In addition, spectral CT can also reduce CT artifacts and provide more accurate corrections for emission data. In this review, we discuss the technical basis, integration pathways, and translational opportunities for combining spectral CT with molecular imaging. We review sequential and integrated PET/DECT approaches, emerging PET/PCCT concepts, PET-enabled spectral CT, and extensions to SPECT/CT. We then examine how spectral CT information may improve attenuation correction, scatter correction, tissue-fraction correction, positron range correction, reconstruction priors, and dosimetry. Clinical and translational applications include contrast-enhanced molecular imaging, vascular and perfusion imaging, quantitative bone marrow imaging, musculoskeletal imaging, and theranostics. Overall, spectral CT integration may move hybrid molecular imaging toward tissue-composition-informed molecular imaging, in which radiotracer signals are interpreted within their material context.**

## I. INTRODUCTION

The integration of x-ray computed tomography (CT) with positron emission tomography (PET) and single-photon emission computed tomography (SPECT) has been one of the most successful developments in clinical molecular imaging. In PET/CT [1], [2] and SPECT/CT [3], CT provides the anatomical reference needed to localize radiotracer uptake, define structural context, and support clinical interpretation. CT also provides attenuation information for quantitative correction of emission data [4], [5]. These roles have made CT an essential component of modern molecular imaging systems [6].

In current clinical systems, the CT scan is commonly acquired using a single x-ray energy spectrum with energy-integrating detectors. This approach is practical and sufficient for many current applications. However, it provides limited ability to characterize tissue composition quantitatively. A CT number in a voxel of a reconstructed image reflects a complex combination of the spatially-varying material density, the elemental composition (which includes energy-dependent attenuation), and the x-ray tube spectra [7]. As a result, different materials may produce similar CT numbers, while the same CT number may correspond to different attenuation properties at the photon energies relevant to molecular imaging [8]. This ambiguity between tissue density and composition is particularly critical in regions containing bone, iodine contrast, metal implants, or heterogeneous tissue compartments [8].

Spectral CT provides a pathway to address this limitation. By acquiring or deriving attenuation information at more than one energy, spectral CT methods can provide information about material composition that is not available from single-energy CT [7], [9]. Dual-energy CT (DECT) and photon-counting CT (PCCT) support a range of outputs, including virtual monoenergetic images, virtual non-contrast images, iodine or calcium maps, effective atomic number maps, electron-density-related information, and multi-material decomposition [7], [10], [11]. These capabilities have been developed for diagnostic CT purposes, but they can also support hybrid imaging with PET and SPECT. Emission tomography measures radiotracer distribution within tissues, for which tissue composition may influence quantification, correction, and biological interpretation.

The added value of spectral relative to standard CT in emission tomography begins with attenuation correction [4], [5], [12]. In PET/CT, CT values must be transformed into attenuation coefficients at 511 keV for attenuation correction [4]. In SPECT/CT, CT-based attenuation correction is similarly important [12], although the correction problem depends on radionuclide energy and system-specific factors [13]. With single-energy CT, this transformation is inherently imperfect because diagnostic x-ray attenuation and emission-photon attenuation depend differently on material composition [8]. The resulting errors may be modest in many soft-tissue regions, but they can become more consequential in the presence of iodine contrast, bone, metal, or other high-atomic-number materials [8], [14], [15], [16], [17]. Spectral CT can reduce this ambiguity by distinguishing material classes to provide more accurate attenuation maps at the needed energies [8].

---

Guobao Wang is with the EXPLORER Molecular Imaging Center, Department of Radiology, University of California Davis Health, Sacramento, CA 95817, United States (email: gbwang@health.ucdavis.edu).

Kris Thielemans is with the Institute of Nuclear Medicine, Respiratory Medicine and the UCL Hawkes Institute, University College London, NW1 2BU, United Kingdom (email: k.thielemans@ucl.ac.uk).

Peter B Noël and Joel S Karp are with the Department of Radiology, University of Pennsylvania, Philadelphia, PA 19104, United States (email: peter.noel@pennmedicine.upenn.edu, joelkarp@pennmedicine.upenn.edu).

Paul E Kinahan is with the Department of Radiology, University of Washington, Seattle, WA 98195, United States (email: kinahan@uw.edu).

Beyond attenuation correction, spectral CT can also improve how molecular imaging signals are interpreted. At the voxel level, emission signals can be confounded by heterogeneous sub-voxel tissue structures, such as trabecular bone and active marrow in bone marrow imaging [18], and air, blood and tissue in lung imaging [19], [20]. Without prior knowledge of tissue composition, the measured radiotracer concentration may reflect both molecular activity and tissue mixture effects. This issue becomes increasingly important as molecular imaging moves toward more quantitative applications, including kinetic modeling, therapy-response assessment, and radiopharmaceutical dosimetry [21]. In this view, CT is no longer only a supporting modality for anatomical localization and attenuation correction. It becomes an active quantitative method that characterizes the material environment in which radiotracer uptake occurs. This integration may enable a broader class of tissue-composition-informed molecular imaging, in which emission tomography measures biological processes while spectral CT characterizes the physical and compositional context of those processes.

Several technological developments make this opportunity timely. DECT is now clinically available through multiple implementations, including sequential acquisition, rapid tube-potential switching, dual-source CT, and dual-layer detector CT approaches [7]. PCCT is emerging as a more general spectral CT platform with energy-resolved detection, improved spatial resolution, reduced radiation dose, and potential for K-edge imaging [10], [22]. In parallel, emission tomography is advancing toward more quantitative applications, supported by advanced image reconstruction methods [23] and growing interest in quantitative tracer kinetics [21], [24], [25] and theranostics [26], [27]. In PET, this trend is further accelerated by improved time-of-flight detectors [28], [29] and total-body and long axial field-of-view (LAFOV) systems [30], [31], while SPECT continues to benefit from advances in semiconductor detectors and collimator design [32], [33], [34]. These parallel developments create an opportunity to integrate material information from spectral CT with molecular, functional, and kinetic information from emission tomography.

The purpose of this manuscript is to articulate this emerging opportunity and discuss the technical and translational aspects of integrating spectral CT with molecular imaging. We first summarize the relevant spectral CT imaging methods and then describe major integration pathways, followed by possible technical applications. Finally, we discuss clinical and translational examples in which tissue-composition information may improve the interpretation and quantification of measurements from emission tomography. By organizing these developments around the concept of tissue-composition-informed molecular imaging, this review paper aims to provide a perspective for future research, system development, and clinical translation.

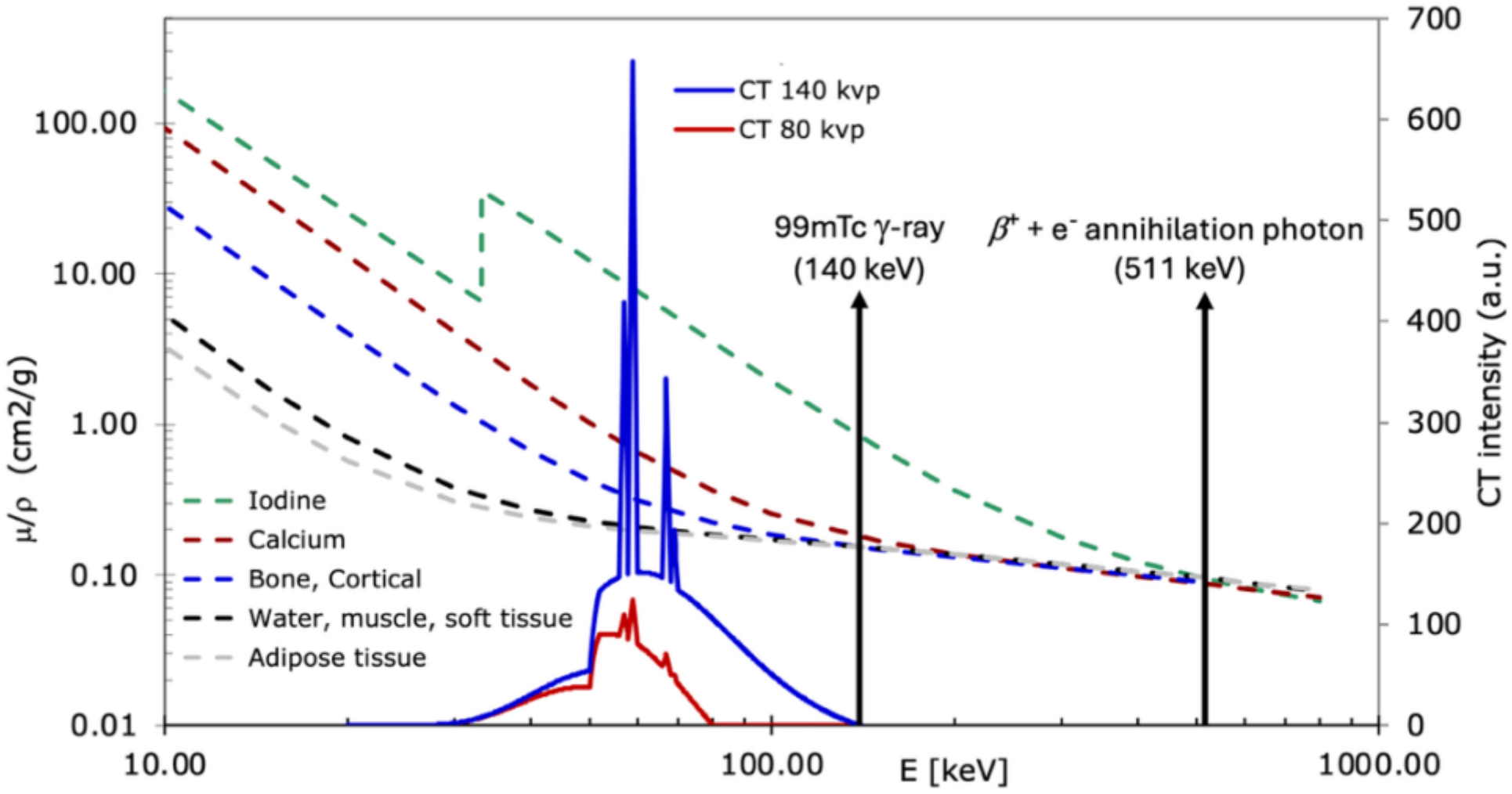

Figure 1. Energy dependence of photon mass-attenuation coefficients for representative materials. Also shown are typical CT x-ray tube spectra for 80 and 140 kVp and two of the most common energies for SPECT and PET.

## II. Spectral CT Imaging Methods

### *A. Dual-Energy CT Acquisition Methods*

DECT is the most widely used form of spectral CT. DECT obtains two energy-dependent measurements, typically from low- and high-energy x-ray spectra, and exploits differences in energy-dependent attenuation to separate materials [7] (Fig 1). Several acquisition strategies have been developed, which differ in how they trade off spectral separation, temporal and spatial registration of the two energy measurements, and hardware complexity.

The first, and simplest approach is temporally sequential DECT, in which two CT scans are acquired consecutively at different tube potentials. This approach can be implemented on many conventional scanners without dedicated dual-energy hardware, and the tube current and spectral filtration can be optimized independently for each scan. Its major limitation is the temporal gap between the two acquisitions: patient body motion, respiratory motion, or other organ motion can introduce misregistration between the low- and high-energy images, degrading material decomposition [7].

A second approach is rapid tube-potential switching, in which the x-ray tube alternates between low and high tube potentials from projection to projection during a single acquisition [11], [35]. Because consecutive low- and high-energy projections are acquired at nearly the same angular position, the two data sets are approximately registered in the projection domain, enabling projection-domain as well as image-domain material decomposition and providing good temporal consistency. Its limitations include the need for a specialized high-frequency generator and fast detector readout, constraints on tube-current modulation (only partially addressed on newer-generation systems), and comparatively limited spectral separation, because the two beams cannot be filtered independently and the finite rise and fall times of the tube potential further blend the spectra [7].

A third approach is dual-source DECT, in which two x-ray tube-detector pairs, mounted with an angular offset of approximately 90°, acquire data simultaneously at different tube potentials [36], [37]. This design allows the tube current and spectral filtration to be optimized separately for each spectrum (for example, added tin filtration of the high-energy beam) yielding the best spectral separation among source-based techniques. Acquisition of the two data sets is near-simultaneous, offset by roughly a quarter of a gantry rotation, which makes the approach robust for moving anatomy and contrast-enhanced CT imaging. However, because the low- and high-energy projections are acquired at different angular positions, they are not geometrically consistent, and material decomposition is therefore generally restricted to the image domain. Additional limitations include specialized hardware requirements, a reduced dual-energy field of view for the second detector, cross-scatter between the two source–detector pairs, and the need for careful calibration and correction.

A fourth approach is split-filter DECT (marketed as TwinBeam Dual Energy), in which the beam of a single x-ray source operated at a fixed tube potential is divided along the z-axis by two different filter materials (typically gold and tin) producing adjacent low- and high-energy beam halves that irradiate the patient during the same helical acquisition [38]. Its principal advantage is that dual-energy data can be obtained on a single-source scanner with largely conventional hardware and at standard radiation dose levels. Its limitations include the weakest spectral separation of the source-based approaches, since both spectra originate from the same tube potential; a temporal and z-positional offset between the two beam halves at each slice location, which reintroduces some sensitivity to motion; reduced dose efficiency due to the added filtration; and restrictions on helical pitch, with decomposition performed in the image domain [38].

A fifth approach is dual-layer (or multilayer) detector CT [39], in which a single x-ray source is used and the energy separation is performed at the detector: an upper detector layer preferentially absorbs lower-energy photons while a lower layer absorbs the remaining higher-energy photons [11], [40]. Because both energy measurements are obtained from the same ray at the same instant, temporal and spatial misregistration is essentially eliminated, enabling projection-domain material decomposition; moreover, spectral information is available retrospectively from every routine acquisition, simplifying workflow. Its limitations include substantial overlap between the effective energy spectra, which is fixed by the detector design and cannot be tuned per protocol. The fixed spectral overlap must therefore be accounted for in reconstruction and material decomposition [39], [40].

### *B. Radiation Dose of DECT*

Radiation dose considerations differ across these DECT implementations, although none inherently requires a dose increase over single-energy CT [41]. In temporally sequential DECT, the tube current-time product of each scan must be reduced so that the combined dose of the two acquisitions remains comparable to a single-energy examination, with dose partitioning between the low- and high-kV scans optimized to balance noise. Rapid tube-potential switching delivers the total dose within a single acquisition at levels generally comparable to single-energy CT, although the historical constraints on tube-current modulation could raise dose for some patient sizes and protocols relative to a modulated single-energy scan. Dual-source DECT is generally the most dose-efficient source-based implementation: the tin filtration of the high-energy beam removes low-energy photons that contribute dose but little spectral information, tube-current modulation is available independently on both tubes, and dose-neutral operation relative to single-energy protocols has been demonstrated, with cross-scatter between the two source-detector pairs only modestly reducing dose efficiency. In split-filter DECT, the gold and tin filters attenuate the beam before it reaches the patient, so patient dose remains at routine levels at the cost of increased tube loading; however, the weak spectral separation makes material-specific results noisier at matched dose, lowering the effective dose efficiency for spectral tasks. Dual-layer detector CT is dose-neutral by construction, as the acquisition is identical to a conventional scan and all standard dose-reduction tools apply unchanged; its main penalty is likewise one of efficiency, since the fixed spectral separation of the layered detector increases noise in material-decomposed images at a given dose.

### *C. Photon-Counting CT and Multi-Energy Imaging*

PCCT represents a broader multi-energy approach [22], [42]. Unlike conventional energy-integrating detectors, photon-counting detectors directly convert individual x-ray photons into electrical signals, count photons, and sort them according to energy thresholds [43]. This approach can reduce the influence of electronic noise, improve spatial resolution, and provide energy-resolved measurements from a single acquisition. PCCT can generate spectral reconstructions similar to DECT, including virtual monoenergetic images, virtual non-contrast images, iodine maps, effective atomic number maps, and multiparametric images [10], [44], [45]. These detector characteristics may also improve radiation-dose efficiency, allowing comparable image quality at lower dose in some applications; however, the achievable dose reduction depends on detector design, patient size, acquisition protocol, reconstruction method, and imaging task.

Because PCCT can use multiple energy bins, it may also support K-edge imaging for selected contrast materials [10], [46]. This capability may become important for molecular imaging applications, particularly if new contrast agents are designed to exploit K-edge separation. At the same time, PCCT has technical challenges, including pulse pile-up, charge sharing, detector calibration, count-rate limitations, and material-decomposition noise [22], [44]. These limitations are important because integration with emission imaging will require not only high-quality diagnostic CT images, but also quantitatively reliable material information.

For hybrid molecular imaging, PCCT is attractive because it may provide improved anatomical detail, potentially better separation of high-atomic-number materials, reduced blooming artifacts from calcification or metal, and richer material information for quantitative modeling [10], [22], [40], [45]. However, PCCT is not necessarily an ideal choice for integration simply because it improves CT. Its value for molecular imaging will depend on whether the additional spatial and spectral information improves emission-image correction, quantification, reconstruction, or clinical interpretation in specific applications.

### *D. Material Decomposition as the Bridge to Molecular Imaging*

Among DECT and PCCT, the key computational output is material decomposition [44], [47]. Energy-dependent attenuation measurements can be represented using physical interaction bases, such as photoelectric absorption and Compton scattering, or material bases, such as water, iodine, calcium, bone, fat, or air [43]. Depending on the acquisition method and implementation, material decomposition may be performed in the projection or image domain using analytic, statistical, or iterative methods [7], [43], [48].

In the following sections, material decomposition is shown to be important as the bridge between spectral CT and molecular imaging. Material-specific and derived images—including iodine maps, calcium or bone fraction maps, electron-density-related images, effective atomic number images, and virtual monoenergetic images—can provide physical information complementary to radiotracer measurements. Their specific roles in attenuation correction, scatter correction, tissue-fraction correction, reconstruction priors, positron range correction, dosimetry, and clinical interpretation are discussed in Section IV and Section V.

## III. Integration Pathways Between Spectral CT and Emission Tomography

The integration of spectral CT with emission tomography can occur through several pathways, ranging from sequential imaging workflows to fully integrated hardware systems. Spectral CT can complement both PET and SPECT, although the limited number of integrated solutions and studies reported to date have focused primarily on PET. Accordingly, the pathways and examples in this section use PET as the primary example (Table 1), while SPECT-specific considerations are discussed separately at the end of the section.

TABLE 1.
Integration Pathways Between Spectral CT and PET

| Integration pathway | Basic Concept | Main Advantages | Main Limitations | Relevant Applications |
|---|---|---|---|---|
| Sequential PET + DECT/PCCT | Separate PET and spectral CT exams analyzed together | Easy research entry point; no new hybrid hardware required | Registration, timing mismatch, workflow, possible extra dose | Early validation; vascular/perfusion; tissue-composition studies |
| Integrated PET/DECT | PET/CT system with DECT-capable CT | Same-session acquisition; improved alignment; near-term clinical route | Hardware availability; DECT implementation tradeoffs; dose/noise balance | Contrast-enhanced PET/CT; attenuation correction; bone / marrow studies |
| PET/PCCT | PET combined with photon-counting CT, sequential or integrated | Richer material information; high spatial resolution; potentially improved dose efficiency; K-edge imaging | Emerging; PET-specific benefit requires validation | Vascular imaging; bone/calcium; metal artifacts; new contrast agents |
| PET-enabled Spectral CT | Time-of-flight (TOF) PET emission data used to derive 511-keV gamma-ray attenuation combined with x-ray CT | No second x-ray CT scan; may work on existing TOF PET/CT; complementary to DECT/PCCT | Requires TOF PET; early validation stage | Material decomposition; tissue composition |

### A. Sequential Approaches

The most straightforward pathway is sequential imaging, in which PET is combined with a separate DECT or PCCT examination [10], [14]. This approach can be useful in early research studies since it allows investigators to evaluate spectral CT information without requiring a fully integrated PET/spectral CT scanner. In this role, sequential imaging can serve as a translational bridge, allowing PET and spectral CT information to be studied together before dedicated integrated systems or workflows are available.

The limitations of sequential imaging are also clear. Separate acquisitions introduce registration challenges, differences in patient positioning, respiratory or cardiac motion mismatch, longer scan duration, and potentially additional radiation exposure. These limitations become more important when spectral CT data are used quantitatively rather than visually. For example, a bone fraction map, iodine map, or electron-density-related image must be spatially aligned with the PET image if it is used for attenuation correction, voxel-wise tissue-fraction correction, kinetic modeling, or dosimetry. Small mismatches may matter less for broad visual interpretation but can become important when spectral CT images are used as quantitative inputs to PET analysis.

### B. Integrated Hardware Approaches

A more direct pathway is to pair PET with sequential DECT acquisitions [15], [16], [49], [50] or with dedicated integrated PET/DECT hardware [51], [52]. However, commercial adoption has remained limited; one integrated commercial PET/DECT system has been reported in the literature [53]. In this configuration, spectral CT data are acquired during the same imaging session as PET and can be used for anatomical localization, attenuation correction, material decomposition, and diagnostic interpretation. Compared with separate imaging, integrated PET/DECT improves spatial alignment, reduces workflow complexity, and allows the CT acquisition to be coordinated with the PET protocol. In addition, work at the University of Pennsylvania has combined the long axial field-of-view (LAFOV) PennPET Explorer [54] with the dual-layer Philips IQon CT to create an integrated research PET/DECT system [55]. The IQon CT is used routinely for attenuation correction and anatomic localization, as with conventional CT, but is also providing opportunities to leverage spectral CT capabilities for improvement of PET/CT imaging, including blood flow quantification and metal artifact reduction [56], [57], see Fig. 2 for an example.

The specific DECT implementation may matter. Sequential DECT may be feasible but remains sensitive to motion and contrast dynamics. Rapid kV switching and dual-source CT can reduce temporal mismatch, but they require specific hardware and have different tradeoffs in spectral separation, field of view, noise, calibration, and radiation dose. For PET, these tradeoffs matter because spectral CT images may be used not only for visual diagnosis but also as quantitative inputs to PET correction and modeling (see Section IV for details).

Integrated PET/DECT is particularly relevant for contrast-enhanced PET/CT. Iodinated contrast can improve anatomical and vascular assessment, but it can complicate CT-based attenuation correction when single-energy CT is used [8]. DECT can help separate iodine from bone or soft tissue and can support more appropriate generation of attenuation maps [14], [15]. From a system-design perspective, integrated PET/DECT should not be viewed simply as PET plus a better CT scanner. Its value depends on whether the spectral CT information is used in the PET workflow. The most meaningful integration can occur when DECT-derived tissue-composition maps are incorporated into attenuation correction, reconstruction, kinetic modeling, dosimetry, or interpretation of PET signal in mixed tissues.

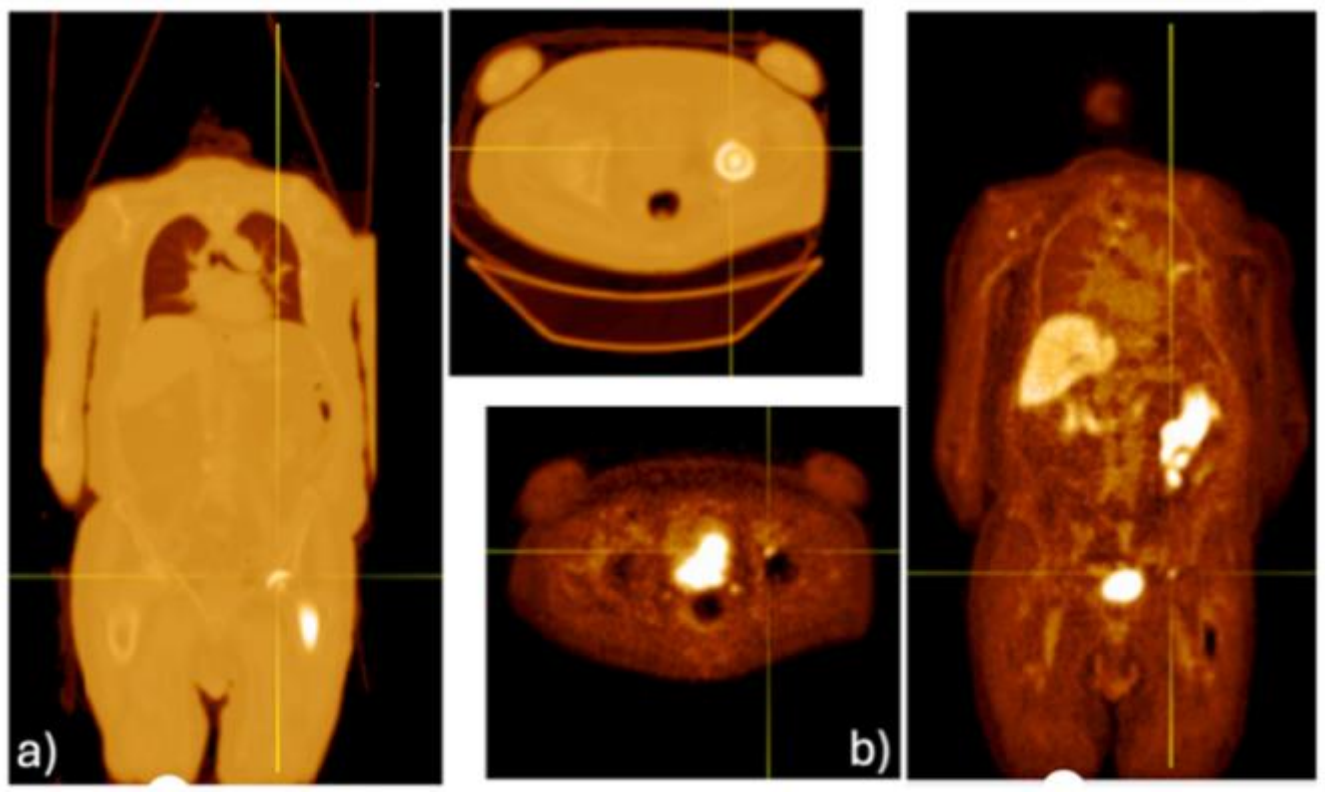

Figure 2: Example of patient acquired on PennPET Explorer / dual-layer spectral CT system with metal hip prosthesis showing a) attenuation image generated from spectral CT and b) CTAC PET image. Measurement of the $^{18}$F-fluorodeoxyglucose (FDG) uptake of the lesion at the edge of the metal implant is impacted by the accuracy of the CTAC and its ability to mitigate the artifacts due to the metal implant.

### C. PET with PCCT

PCCT creates another integration pathway, although a fully integrated PET/PCCT system has not yet been reported. The potential value of PET/PCCT is broader than PET/DECT or PET/CT given the advantages of PCCT compared to DECT or standard CT. PCCT may improve anatomical detail, reduce blooming artifacts from calcification or metal, and has the potential to provide richer material information than conventional CT or DECT. These capabilities may potentially further improve PET attenuation correction, artifact reduction, vascular and perfusion imaging, bone and calcium characterization, tissue-fraction correction, and anatomical or material priors for PET reconstruction.

The design requirements for a PET/PCCT system will depend strongly on the intended application. For example, a PET/PCCT workflow designed primarily for attenuation correction may require accurate and robust generation of 511-keV attenuation maps. A workflow designed for vascular imaging may require iodine quantification, contrast timing, and motion management. An application designed for theranostics may require material-density information for organ mass estimation and absorbed-dose calculation. These use cases may place different demands on spectral resolution, spatial resolution, scan timing, dose, and calibration.

PET/PCCT may also have an important role if new contrast materials are developed for K-edge imaging or if CT-derived material maps become part of multiparametric molecular imaging protocols [10], [58]. A future PET/PCCT workflow could combine radiotracer uptake with iodine maps, calcium maps, effective atomic number images, or K-edge contrast-agent maps. At present, these directions remain emerging opportunities and require direct validation in PET-centered studies.

*D. PET-Enabled Spectral CT*

A different pathway is the use of 'PET-enabled spectral CT'. Instead of requiring a second x-ray CT with different spectral information, PET-enabled DECT [59], [60] uses the time-of-flight (TOF) PET emission data to reconstruct a 511-keV gamma-ray attenuation image and combines this high-energy attenuation image with the conventional x-ray CT image already acquired in the PET/CT scan (Fig 3). The 511-keV gamma-ray attenuation image can be estimated directly from TOF PET emission data using the maximum-likelihood attenuation and activity (MLAA) type algorithms [61]. Without TOF information, the estimation becomes more difficult but remains possible if multiple energy windows are available for PET data [62].

The key feature of PET-enabled DECT is that it creates a dual-energy pair using two different physical sources: a lower-energy x-ray CT image and a higher-energy 511-keV gamma-ray attenuation image. This makes the method conceptually different from conventional x-ray DECT, in which both measurements are x-ray based. It also makes the method complementary to hardware-based DECT and PCCT systems, rather than a replacement for them. PET-enabled DECT may provide a route to material decomposition on existing time-of-flight PET/CT systems without additional x-ray dose or hardware replacement. It can be also combined with x-ray DECT for tri-energy CT imaging.

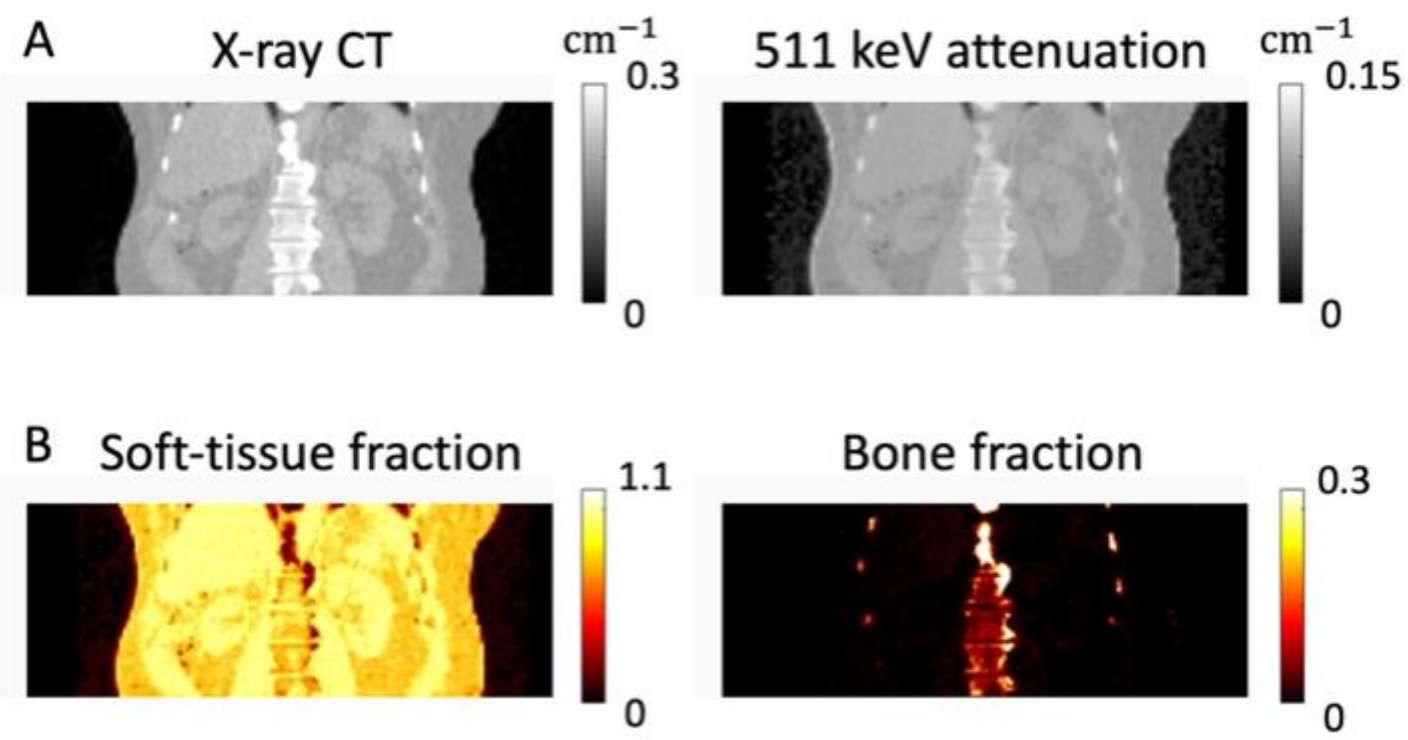

Fig. 3: Example of PET-enabled dual-energy CT imaging. (A) a lower-energy x-ray CT is paired with a higher-energy 511 keV attenuation image derived from a time-of-flight PET/CT scan of a human subject. (B) Result of material decomposition for soft tissue and bone fractions. Shown are a coronal plane.

The main technical challenge is that the gamma-ray 511-keV attenuation image must be reconstructed from PET emission data, which are noisy [59], [63] if standard MLAA reconstruction is used. The 'Kernel MLAA' method, inspired by the kernel methods in machine learning, addresses this issue by using the available x-ray CT image as prior information to guide gamma-ray attenuation image reconstruction [59], [63], [64]. This formulation uses the anatomical information in x-ray CT while still estimating the 511-keV attenuation information from PET emission data.

Recent work has demonstrated the feasibility of PET-enabled DECT in physical phantom and initial patient studies [60]. The current results are encouraging. Further work is needed to evaluate robustness across scanners, tracers, motion conditions, and clinical applications.

*E. Extension to SPECT/CT*

Although the preceding discussion focuses primarily on PET, the same considerations apply to SPECT combined with spectral CT. As for PET, CT can provide anatomical localization and physical information for attenuation and scatter correction in SPECT imaging. Spectral CT can extend this role by providing material information that can improve quantification, dosimetry, and interpretation. However, the technical requirements for SPECT are not identical to those for PET.

One major difference is photon energy. PET attenuation correction is centered on 511-keV annihilation photons, whereas SPECT uses radionuclide-specific photon energies [3], [27], such as those from $^{99m}$Tc, $^{123}$I, $^{111}$In, $^{177}$Lu, or other therapeutic and diagnostic radionuclides. Spectral CT-derived material maps may therefore need to support attenuation-map generation across multiple photon energies rather than a single PET energy. This requirement may be especially relevant for quantitative SPECT and theranostic imaging, where absorbed-dose estimation depends on accurate activity quantification and tissue-specific attenuation correction.

In SPECT, attenuation depends on the depth and direction of photon emission relative to the detector, making attenuation correction strongly projection-dependent. CT-based attenuation maps can greatly simplify this problem [65]. SPECT also has additional system-dependent factors [3], [13], including collimator-detector response, septal penetration, scatter, partial-volume effects, and energy-windowing. Spectral CT can help address some of these issues by providing material and density information for attenuation and scatter modeling, but the benefit will depend on the photon energies of the radionuclide, acquisition protocol, reconstruction method, and calibration. Compared with PET, the integration of spectral CT with SPECT may therefore require more radionuclide-specific validation. On the other hand, the presence of multiple photon energies from a single radionuclide or its daughters could enable extension of PET-enabled spectral CT to SPECT.

The clinical motivation may be particularly strong in theranostics. In targeted radionuclide therapy, quantitative SPECT/CT is increasingly used for patient-specific dosimetry [27]. Spectral CT-derived material information could support organ and tissue mass estimation, bone and marrow separation, metal artifact reduction, and potentially tissue-specific modelling of transport of $\alpha$, $\beta$ and $\gamma$ radiation. These capabilities may be important for bone marrow dosimetry, lesion dosimetry, and treatment planning, especially when heterogeneous tissues or high-density materials are present.

### *F. Cost and Value Considerations*

Cost is relevant to all spectral CT integration approaches because high-end spectral CT systems (e.g., PCCT) require substantially greater capital investment than the conventional CT components used in many PET/CT systems. This is even more the case for SPECT. The key question, however, is whether PCCT provides sufficient incremental value when combined with emission tomography. This value may arise from improved correction and quantification, replacement of separate diagnostic CT examinations, more informative characterization of tissue composition, streamlined workflows, or new multiparametric applications.

Total-body (or LAFOV) PET [66], [67], [68], [69] provides a useful precedent. Despite its high capital cost, it has been transformational for both clinical and research applications by enabling qualitatively new capabilities in sensitivity, whole-body dynamic imaging, and low-dose studies. However, the rationale for investing in a total-body/LAFOV PET system depends on whether it improves clinical workflow [70] or enables holistic imaging of multi-organ systems in the body that is not otherwise possible with standard PET systems [71]. Similarly, the long-term value of PET/spectral CT will depend on whether it enables clinically or scientifically transformative applications rather than incremental improvements in CT image quality alone. Cost-effectiveness should therefore be evaluated on an application-specific basis, considering both the added investment and the downstream clinical, operational, and scientific benefits.

## IV. Technological Applications of Spectral CT to Improve Molecular Imaging

The technical value of spectral CT for molecular imaging begins with material decomposition. In addition to providing a conventional CT image, spectral CT can generate material-specific or interaction-specific information that is not available from single-energy CT [7]. For hybrid molecular imaging, this information creates a bridge between CT physics and emission-imaging physics, enabling improvements in emission-image quantification. This section focuses on technical uses of spectral CT information in molecular imaging (Table 2). Material maps can potentially support more accurate attenuation-map generation, improve modeling of photon interactions, estimate tissue fractions, guide reconstruction, and provide material context for interpreting radiotracer uptake in heterogeneous tissues. These applications are not equally mature. DECT-based attenuation correction has a longer technical history in PET/CT [8], [14], [15]. Tissue-composition correction for PET quantification is emerging, with early evidence in bone marrow imaging [18]. Other uses, such as spectral CT-informed scatter correction, positron range correction, material-prior reconstruction, and spectral CT-informed SPECT dosimetry, remain promising but require further validation.

TABLE 2.
Spectral CT Information and Molecular Imaging Uses

| Spectral CT Output | Contribution to Molecular Imaging | Illustrative Application |
|---|---|---|
| Material decomposition | Tissue-composition mapping; material-informed reconstruction priors | Bone marrow tissue-fraction correction; lung air-tissue correction; material-informed emission-image reconstruction |
| Iodine maps | Iodine quantification; distinction of iodine from bone or soft tissue; support for attenuation correction | Contrast-enhanced oncologic PET/CT; vascular and perfusion assessment |
| Bone/calcium fraction | Tissue-fraction correction; characterization of osseous composition | Quantitative bone marrow PET; bone lesion and musculoskeletal imaging |
| Virtual monoenergetic images | Artifact reduction; optimization of CT contrast | Molecular imaging near metal implants; interpretation of contrast-enhanced studies |
| Electron-density-related maps | Energy-specific attenuation-map generation; tissue-density or mass estimation | Attenuation and scatter correction; molecular radiotherapy dosimetry |
| Effective atomic number | Material characterization and differentiation of high-atomic-number materials | Differentiation of iodine, calcium, and metal; plaque or calcification assessment; molecular radiotherapy dosimetry |

### *A. Attenuation Correction*

Attenuation correction is an established technical motivation for integrating spectral CT with emission tomography [4], [5], [8], [12]. In PET/CT, the CT image is transformed into an attenuation map at 511 keV [8]. In SPECT/CT, the target photon energy depends on the radionuclide(s) [3], [72]. In both settings, the accuracy of the attenuation map will affect quantitative activity estimation.

Single-energy CT-based attenuation correction relies on mapping CT numbers acquired at diagnostic x-ray energies to attenuation coefficients at emission photon energies. This mapping is imperfect because x-ray attenuation and gamma-ray attenuation depend differently on material composition [8]. The ambiguity is particularly important for bone, iodine contrast, calcium, metal, and other high-atomic-number materials. In PET/CT, iodine contrast can be incorrectly treated as bone or otherwise mis-scaled when single-energy CT is transformed into a 511-keV attenuation map [8], [15]. These errors may not always alter visual interpretation, but they can affect quantitative PET measures when accurate uptake estimation is required [8].

DECT can reduce this ambiguity by providing energy-dependent information that helps distinguish material classes. Early work on DECT attenuation correction showed that dual-energy information can improve estimation of PET attenuation maps by separating components such as Compton scatter and photoelectric absorption or by distinguishing materials such as iodine and bone [8], [50]. Later studies further explored DECT-based approaches for contrast-enhanced PET/CT attenuation correction and electron-density-based attenuation-map generation [14], [15]. These approaches are attractive because they address a limitation of single-energy CT that is directly linked to PET quantification.

Several practical issues remain important. DECT attenuation correction may reduce material-scaling bias but may, depending on the implementation, introduce challenges related to noise propagation and, for sequential acquisitions, image registration [7], [48]. Material-decomposition noise may propagate into the synthesized attenuation map. Low-dose DECT protocols are highly desirable to minimize patient radiation exposure in hybrid imaging, but the reduced photon statistics can substantially degrade the precision of material decomposition. This limitation can be managed by incorporating advanced algorithms [48], [73], [74]. Consequently, low-dose DECT-based attenuation correction should be evaluated not only by CT image quality, but also by its effect on reconstructed activity images [14], [50].

The same principle applies to SPECT, though the implementation is more radionuclide-specific. For radionuclides such as $^{99m}$Tc, the smaller energy difference between x-ray CT and the 140-keV photopeak may reduce material-scaling errors relative to 511-keV PET. However, SPECT acquisitions may use multiple photopeak and scatter energy windows, for which spectral CT-derived material maps may support energy-dependent attenuation-map and scatter modeling, particularly in regions with bone, metal, contrast agents, or heterogeneous tissue composition. The value of spectral CT for SPECT attenuation correction will need to be validated for specific radionuclides, energy windows, reconstruction methods, and dosimetry tasks.

### *B. Scatter Correction*

Scatter correction is another potential area where spectral CT may contribute to quantitative emission imaging. Scatter estimation depends on object geometry and photon-interaction properties, commonly modeled using attenuation information [75], [76]. In current emission-imaging workflows, scatter correction is usually performed using system models, energy-window methods, simulation-based approaches, or approximations based on single-energy CT-derived attenuation maps. Spectral CT may improve these models by providing more accurate material and density information, especially in anatomically heterogeneous regions or in the presence of high-density materials.

The evidence for spectral CT-enabled scatter correction is less mature than for attenuation correction. A conservative framing is therefore appropriate. Spectral CT-derived electron-density-related images, effective atomic number maps, or material maps may improve the physical modeling of scatter, but this needs direct validation in emission imaging. For SPECT, where scatter can be strongly radionuclide- and energy-window-dependent [3], material information may be useful when integrated with collimator-detector response and radionuclide-specific reconstruction models.

### C. Tissue-Fraction Correction

Tissue-fraction correction is a direct example of how spectral CT can move hybrid molecular imaging beyond attenuation correction. While emission tomography captures the net radiotracer concentration within an image voxel, individual voxels may contain multiple discrete tissue compartments. Spectral CT-derived material maps can help estimate the fractional composition of the voxel and thereby support more accurate interpretation of the emission signal.

Bone marrow imaging provides a representative example. A PET-defined "bone marrow" region may include marrow tissue and trabecular bone. DECT material decomposition can estimate voxel-wise bone volume fraction (see Fig. 3B for example) and use this information to correct PET standardized uptake value (SUV) and kinetic parameters [18]. This illustrates that spectral CT can help define the tissue compartment in which radiotracer concentration is interpreted, beyond its conventional role in attenuation correction.

For SPECT, tissue-fraction information may also be valuable in personalized dosimetry, see Section V.D.

### D. Positron Range Correction

Positron range correction is a PET-specific application in which CT-derived tissue information can be important [77]. Before annihilation, emitted positrons travel a finite distance that depends on positron energy and the surrounding material [78]. This effect causes spatial blurring in PET, especially for high-energy positron emitters [79], [80] such as $^{68}$Ga, $^{82}$Rb, $^{13}$N, and $^{15}$O. The positron range is tissue-dependent. Positrons travel farther in low-density tissues such as the lung than in soft tissue or bone, with its effect non-intuitive in non-homogeneous media [81], [82].

CT-derived attenuation or density information has been used to estimate spatially varying positron range kernels [77], [83], as well as for deep-learning based approaches [84], [85]. However, positron range depends on many material properties, including electron density and atomic number [86]. Spectral CT may improve this direction by providing more accurate tissue classification or material composition than single-energy CT. For example, material maps may help distinguish lung, soft tissue, trabecular and cortical bone, and other structures more robustly for positron range modeling. This may be important near tissue interfaces, where positron range blurring can mis-allocate activity across anatomical boundaries [81], [82]. This application can be most relevant for PET tracers with higher positron energies and for imaging tasks where boundary accuracy or small-structure quantification is important.

### E. Anatomical and Material Priors for Image Reconstruction

Spectral CT may also contribute to emission-image reconstruction by providing anatomical and material priors. Anatomical priors from co-registered images [87] have long been explored (e.g., [88], [89], [90]) and remain active (e.g., [23], [91], [92], [93]) to improve emission tomography reconstruction, reduce noise, and preserve edges. Spectral CT can extend this concept by providing not only anatomy but also material information. A bone map, iodine map, calcium map, air-tissue map, or effective atomic number image may provide more task-relevant prior information than a conventional CT image alone.

Material-informed priors for emission reconstruction may be useful in several settings. For example, bone or calcium maps may help guide image reconstruction near osseous structures or improve tissue-fraction correction after reconstruction. Iodine maps may help distinguish contrast-enhanced blood pool or vascular structures from radiotracer uptake. Air-tissue maps may support reconstruction or partial-volume correction in the lung.

The image priors may be particularly relevant for low-count PET, theranostic imaging, and SPECT, where noise can limit quantitative reconstruction. Material information could be incorporated through regularization, kernel methods, deep neural networks (e.g., [23], [94], [95]). The appropriate use of spectral CT priors may be task-specific: the image prior should constrain physical or material structure when that information is relevant, but should not force the emission image to reproduce CT contrast that has no biological relationship to tracer uptake.

## V. Clinical and Translational Applications

One motivation for integrating spectral CT with emission tomography is to bring tissue-composition information into molecular imaging. The following examples illustrate clinical and translational applications in which material information may change the correction, quantification, or interpretation of radiotracer measurements.

### A. Contrast-Enhanced Molecular Imaging

Contrast-enhanced PET/CT is clinically attractive because iodinated contrast improves anatomical delineation, vascular assessment, and lesion characterization. For example, in oncology, contrast-enhanced CT may improve the interpretation of PET findings by providing complementary information about lesion location, vascular involvement, tissue enhancement, and disease extent [52], [96], [97]. However, iodinated contrast can complicate CT-based attenuation correction when single-energy CT is used as described above.

Spectral CT provides a direct way to address this problem. DECT can help distinguish iodine from bone or soft tissue, estimate iodine concentration, and support more appropriate generation of attenuation maps (Section IV.A). This would allow contrast-enhanced PET/CT to be used more confidently in quantitative imaging, where SUV, kinetic parameters, or therapy-response metrics are important. The same broad concept may apply to SPECT/CT with radionuclide-specific photon energies and SPECT reconstruction factors taken into account [13].

Beyond attenuation correction, iodine maps from spectral CT may provide complementary vascular or perfusion-related information [10]. When interpreted alongside radiotracer uptake, these maps may provide complementary information about tumor enhancement and treatment response [96]. This direction is clinically appealing, though its value should be tested by application-specific studies rather than assumed from diagnostic CT performance alone.

*B. Perfusion, Vascular Imaging, and Multiparametric Interpretation*

Spectral CT can provide iodine-based measures that are often interpreted as surrogates of contrast delivery, blood volume, and perfusion-related information [10], [40]. These measurements are not equivalent to similar parameters related to radiotracer uptake, but they may provide complementary information about vascular supply and tissue enhancement. When combined with emission imaging, spectral CT may support multiparametric interpretation of tumors, inflammatory lesions, ischemic tissues, vascular and interstitial lung disease.

In oncology, a combined PET/spectral CT workflow could relate radiotracer uptake to contrast enhancement or iodine concentration [51], [96]. For example, $^{18}$F-FDG PET uptake may reflect glycolytic activity, while iodine enhancement may reflect vascularity or perfusion-related properties [51]. In some lesions, the relationship between uptake and enhancement may provide complementary information about tumor heterogeneity and treatment-related change. However, the biological meaning of this combination will depend on tracer, tumor type, contrast timing, and treatment context.

For the purposes of quantifying blood flow in these scenarios, a flow phantom for multi-modal PET and spectral CT imaging has been constructed and tested with the PennPET Explorer/spectral CT at Penn. The modular multimodal flow phantom has a tissue-mimicking compartment, vascular tubing with an adjustable number of pores and valves, and 3D-printed components to support tunable exchange between blood-pool and tissue compartments and controlled dynamic perfusion imaging with same-session PET and spectral CT. Measurements demonstrated the ability to recapitulate $K_1$ (mL/min/mL) across a range of values through adjustable modifications to the phantom configuration, and enable the development of multi-modal approaches for evaluating tissue perfusion with PET/spectral CT [98].

In vascular disease, spectral CT may provide calcium maps, iodine maps, plaque characterization, or improved visualization of vessel wall and lumen [10], [40], [99]. PET can provide molecular information about inflammation, metabolism, calcification activity, or receptor expression depending on the tracer. The integration of these measurements may be useful for plaque assessment, vascular inflammation, active calcification, or treatment monitoring. This application is promising, but careful validation will be needed because CT material properties and PET tracer uptake reflect related but distinct biological processes.

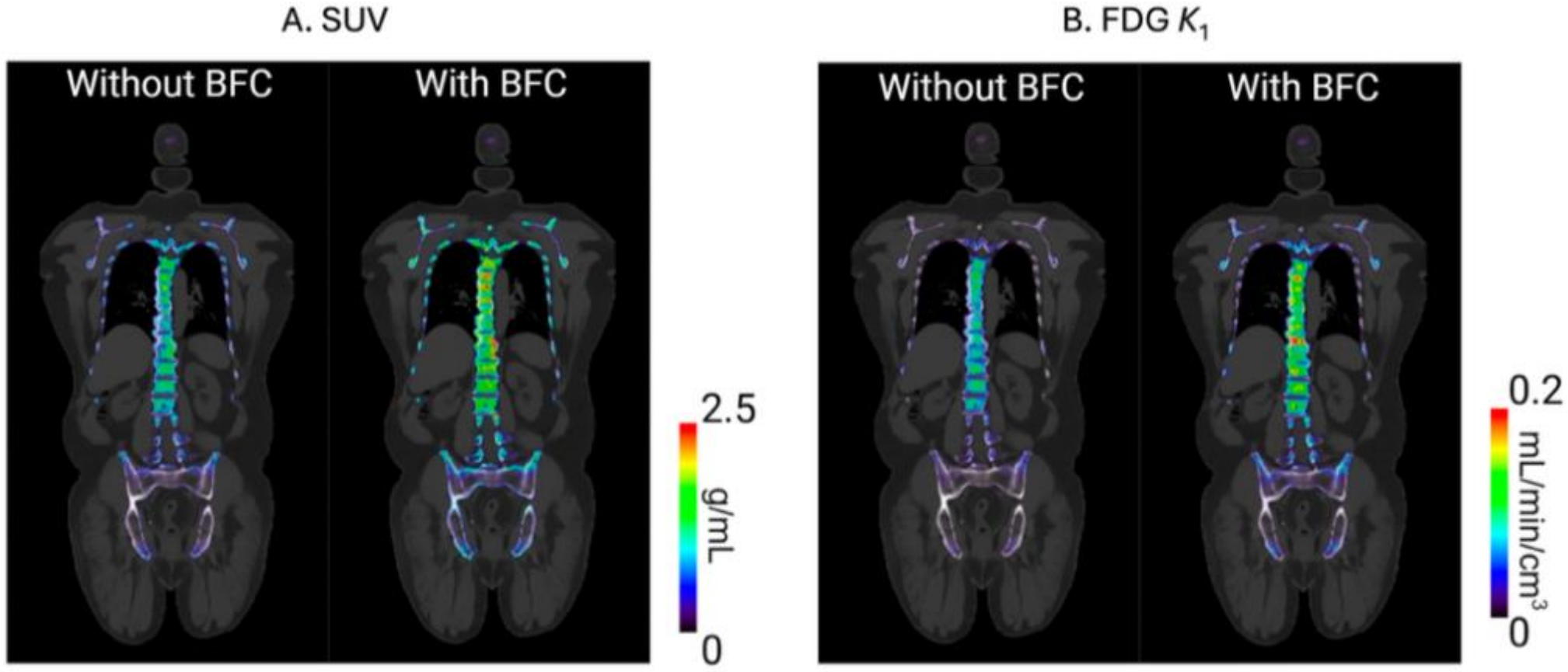

Fig. 4: Example of quantitative bone marrow imaging with $^{18}$F-FDG PET with and without bone fraction correction (BFC) for (A) SUV images and (B) FDG delivery rate $K_1$. The PET images are overlayed on the x-ray CT image. Images are reproduced from the work of Li S. *et al.* [18].

*C. Quantitative Bone Marrow Imaging*

Quantitative bone marrow imaging is clinically relevant in hematologic malignancies, systemic inflammatory states, and so on [100], [101]. The marrow signal is also increasingly accessible with total-body PET, which enables multiparametric assessment of tracer delivery and metabolism in skeletal marrow [101]. However, quantitative interpretation remains challenging because bone marrow regions contain a mixture of active marrow and trabecular bone [18].

Spectral CT provides a way to bring tissue-composition information into bone marrow PET quantification. DECT-derived bone fraction maps can help estimate the marrow tissue fraction within PET-defined skeletal regions and reduce the dilution effect from trabecular bone. Early total-body PET/DECT work [18] showed that incorporating bone fraction information increased $^{18}$F-FDG SUV and kinetic parameter estimates (e.g., FDG delivery rate $K_1$) in bone marrow regions compared with uncorrected measurements (Fig. 4). This application may be useful in diseases or treatments that alter marrow composition, bone density, or marrow metabolic activity over time. Spectral CT-derived bone fraction maps can provide complementary structural and material context for interpreting these PET-based marrow parameters. The clinical value of this approach will require larger studies across disease types, tracers, and treatment settings.

### *D. Theranostics and Personalized Dosimetry*

Tissue composition and in particular positron and electron range affects dosimetry calculations. This could therefore have impact when calculating dose due to radiopharmaceuticals but probably more importantly in theranotics/molecular radiotherapy where the radionuclides are chosen to deliver dose locally, often via $\beta$ emission, with increasing interest in $\alpha$ emitters and Auger-electrons [102], [103]. Current advanced dose calculations use Monte Carlo simulations with CT-derived material decomposition [104], which would therefore become more accurate with access to spectral CT, contributing to patient-specific dosimetry.

Absorbed-dose estimation depends on quantitative activity imaging, time-integrated activity, organ and lesion segmentation, tissue mass and composition, and energy deposition modeling [26], [27], [105]. Single-energy CT already supports anatomical segmentation and attenuation correction in SPECT/CT and PET/CT dosimetry workflows. Spectral CT may extend this role by providing material and tissue-composition information relevant to organ mass, bone and marrow separation, lesion (and other tissue) composition, and artifact correction.

Bone marrow dosimetry is a particularly relevant example [106], [107], [108]. Marrow is often a dose-limiting tissue in radionuclide therapy, but accurate marrow dosimetry is difficult because marrow is distributed within trabecular bone and varies across skeletal regions. Spectral CT-derived bone and marrow fraction information [109] could improve estimates of true marrow volume and tissue mass, which may support more accurate dose calculations when combined with quantitative activity imaging.

Spectral CT may also help in lesion dosimetry, particularly for bone metastases, calcified lesions, heterogeneous tumors, or lesions near metal implants. These applications are promising but remain underdeveloped. Future studies should evaluate whether spectral CT-derived material maps improve absorbed-dose estimates, toxicity prediction, or treatment planning compared with conventional CT-based workflows.

### *E. Musculoskeletal and Bone Applications*

Musculoskeletal and bone imaging provide another potential application. Spectral CT can support virtual noncalcium imaging, bone or calcium mapping, metal artifact reduction, and improved characterization of mineralized tissues [110], [111]. PET can provide complementary information about metabolism, inflammation, bone turnover, or tumor activity depending on the tracer [112], [113], [114].

In oncology, combined PET and spectral CT may be useful for bone metastases and marrow involvement. CT material information can characterize osteoblastic or osteolytic components, while PET can assess tracer uptake related to tumor metabolism, bone remodeling, or treatment response [112], [114], [115]. In non-oncologic musculoskeletal disease [112], [113], combined material and molecular information may be relevant to osteoarthritis, rheumatoid arthritis, infection, prosthesis evaluation, or inflammatory bone disease.

Metal artifact reduction is also clinically relevant in musculoskeletal imaging. Metal hardware can degrade CT images and propagate errors into attenuation correction [16]. Spectral CT approaches may reduce artifacts and improve visualization near implants [44], [110]. For molecular imaging, the endpoint should be whether this improves quantitative emission imaging or clinical interpretation near metal, not simply whether the CT image appears improved.

## VI. Summary and Outlook

Spectral CT integration with molecular imaging is entering a stage where the central question is not only whether spectral CT can provide better CT images, but also whether tissue-composition information can improve molecular imaging correction, quantification, modeling, and interpretation. This requires application-driven development because the spectral CT information needed for attenuation correction may differ from that needed for bone marrow quantification, vascular imaging, or theranostic dosimetry.

Several pathways are likely to develop in parallel. DECT-capable PET/CT systems provide a demonstrated near-term route for incorporating material information into hybrid imaging [52]. PET with PCCT may create a future route toward richer spectral and spatial information, including multi-energy material separation, the potential for lower radiation dose, and K-edge imaging [10], [22]. PET-enabled spectral CT [60] provides an alternative pathway that may extend spectral CT concepts to existing time-of-flight PET/CT systems. For SPECT/CT, spectral CT may be particularly relevant to radionuclide-specific attenuation correction, scatter modeling, and patient-specific dosimetry [3], [27], [65].

Future work may focus on quantitative validation rather than technology demonstration alone. Spectral CT-derived material maps are expected to be evaluated by how they affect emission-imaging endpoints, including SUV, kinetic quantification, absorbed dose, lesion characterization, and clinical decision-making. This will require careful calibration, assessment of noise propagation, motion and registration control, dose optimization, repeatability studies, and multicenter evaluation. The goal should be to define when spectral CT information is necessary, when it is helpful, and when conventional CT is sufficient.

The long-term opportunity is to move hybrid molecular imaging toward tissue-composition-informed molecular imaging. In this framework, PET and SPECT measure biological processes, while spectral CT characterizes the physical and material environment in which those processes occur. Bringing these information layers together may improve quantitative accuracy, reveal tissue-mixture effects, support personalized dosimetry, and enable new multiparametric imaging strategies. Realizing this potential will require collaboration across emission imaging, CT, and clinical application domains. The promise of spectral CT integration with molecular imaging is therefore not simply better hybrid images, but a more complete quantitative description of molecular imaging in its material context.

## Acknowledgment

GBW acknowledges support from NIH grant R01EB036562. KT acknowledges support from CCP SyneRBI, funded by UK EPSRC grant EP/T026693/1, and from STFC through the UKRI Digital Research Infrastructure programme. PBN and JSK acknowledge support from NIH grant R01CA288521. PEK acknowledges support from NIH grant R01CA258298. GBW thanks Dr. Yansong Zhu for assistance in creating Fig. 3 and Dr. Siqi Li for assistance in creating Fig. 4.